# Quantitative exploration of the absorber behavior of metal-insulator-metal metamaterials within terahertz via an asymmetric peak model

**Zhigang Li[1,2], Wenjing Jiang[2], Jianyu Fu[2] and Qing Zhao[1,3], ***

[1]Center for Quantum Technology Research and Key Laboratory of Advanced Optoelectronic Quantum Architecture and Measurements (MOE), School of Physics, Beijing Institute of Technology, Beijing 100081, China
[2]Institute of Microelectronics of The Chines Academy of Sciences, Beijing, 100029, China
[3]Beijing Academy of Quantum Information Sciences, Beijing 100193, China

**Abstract** Terahertz (THz) metamaterials have been developed for THz sensing, detection, imaging, and many other functions due to their unusual absorbers. However, the unusual absorption spectra change with different incident angles. Thus, we designed and fabricated a focal plane array with metal-insulator-metal (MIM) structure metamaterial absorbers for further research. The absorption spectrum with incident angles from 20° to 60° was measured using THz time-domain spectroscopy (THz-TDS), and the experimental results reveal that the absorption spectrum changes with incident angle variations. A basic analytical asymmetric peak model for extracting absorption-frequency characteristics was developed in this study to quantitatively explore this variation in the absorber behavior with incident angles. The best result was that the frequency corresponding to the highest absorption can be easily found using this peak model. The experimental data was coherent with the validation of the asymmetric peak model. Moreover, a second model to quantitatively relate parameters to the incident angle was discovered, allowing for the prediction of absorption spectrum shifts and changes. The absorption spectrum was predicted to have a valley-like absorption curve at particular incident angles based on the secondary model's deduction. The proposed extraction method's essential feature is that it can be applied to any physics-based MIM metamaterial system. Such a model will guide the design and optimization of THz metamaterial absorbers, sensors, imagers, and many others.

## 1 Introduction

Metamaterial absorbers (MMAs) have recently attracted considerable attention and have been widely studied due to their numerous potential applications. The behavior of absorbers can be evaluated using their absorption-frequency characteristics [1-3], which are tested using THz time-domain spectroscopy (THz-TDS) equipment. The absorber's performance depends mainly on the integral geometric characteristics of the structures [4-6] and the incident angle and polarization state [7-9]. This spatial anisotropy of the absorption characteristics makes the design of MMAs more complex. Metamaterials are artificial media with engineered electric permittivities, ε, and magnetic permeabilities, μ, obtained through geometric structuring of natural materials such as metals and insulators [10-12]. In the THz region, MMAs are proposed as a type of metamaterial that plays a key role in perfect absorption [4, 13], sensing [14], imaging [15], thermal emitters [16], and several others. MMAs feature a metal-insulator-metal (MIM) configuration, consisting mainly of a top-patterned metasurface layer, a dielectric layer, and a metal ground layer. Several metasurface layers have been developed based on frequency-selective surfaces [17]. These absorber structures can be demonstrated as a Fabry–Perot (F–P) cavity design. An incident THz wave is resonantly enhanced at given frequencies where the incident electromagnetic energy is trapped within the MMA cavity. For further research on the absorber behavior, several physical modeling, such as equivalent circuits [5, 18, 19], F–P resonant cavities [8], interference theory [20], and transmission lines [21, 22], are usually used to investigate MMAs.

---

[a]e-mail: qzhaoyuping@bit.edu.cn (corresponding author)

Numerical simulation [23-25] is also used to design perfect absorbers. Most of these simulations are based on the finite element method, which can be performed in commercial software, such as CST Microwave studio [5, 26, 27], COMSOL Multiphysics [28, 29], HFSS [4, 30], and FDTD [31, 32]. However, because simulation results are not always consistent with experimental results, an analytical mathematical model for quantitatively exploring the absorption behavior is required. The analytical model can explain the physical mechanism of absorption in greater depth. This study provided a novel concept of the asymmetry model as a function of absorption-frequency characteristics. Therefore, we combined Fermi–Dirac's and Boltzmann's statistics to analyze and obtain the solution satisfying the experiment and the asymmetric model in the terahertz band, which is verified by third-party experimental results.

The secondary model of parameters versus incident angles was found from each parameter coefficient of the model to predict absorber behavior. The absorption characteristics of the absorber changed from peak to valley at a specific incident angle. This model can be applied to the terahertz band and visible, microwave, infrared, and other frequency bands. A model in an explicit functional form to predict the absorption frequency is required, not only by integrating the incidence proceeding to the experimental absorption-frequency curve but also by academic researchers and the division of research manufacturers as emerging topics.

## 2 Experiments

The literature [7, 8, 33] reported the absorption characteristics of MMAs varied with incident angles. D. Yu. Shchegolkov [8] and others reported metamaterial-based absorbers composed of Au/PI/Au structure. The spatial anisotropy of the absorption characteristics was characterized using THz-TDS with the incident angles from 20° to 60°. This is an important phenomenon in semiconductor devices. For further research, we designed and fabricated a square patch metamaterial absorber (SQMA) focal plane array (FPA) based on an Au/SiN$_x$/Al structure. Fig. 1a shows the scheme of a pixel in the SQMA FPA. The pixel size is 140 × 100 μm. The square patch size is 40 × 40 μm, which is the frequency selective surfaces (FSS) structure. Four layers can be observed in the pixel's cross-sectional view from top to bottom: the first layer is used as the FSS structure layer; the second layer is the dielectric layer; the third layer is a metal layer acting as a reflector; the fourth layer is the support structure.

All of the samples were prepared using conventional micro/nanofabrication processes. The fabrication procedure is as follows: a) a 5000 Å thick low-pressure chemical vapor deposition low-stress LPSiN$_x$ layer is generated on two sides of the 4-inch wafer; the stress of SiN$_x$ is easier to adjust and control for large-scale metamaterial [34] fabrication; b) a 1000 Å thick Al layer is evaporated on the front side, and 2.0 μm of Plasma Enhanced Chemical Vapor Deposition PESiN$_x$ layer is deposited on top of the Al layer; c) reactive ion etching is performed to open the back etching window; d) lithography is performed to pattern the FSS structure layer and lift off; e) Silicon wet etching is performed from backside to remove the substrate. There are 4 chips in one wafer, as shown in Fig. 1b. The chip is finally diced by a laser scriber.

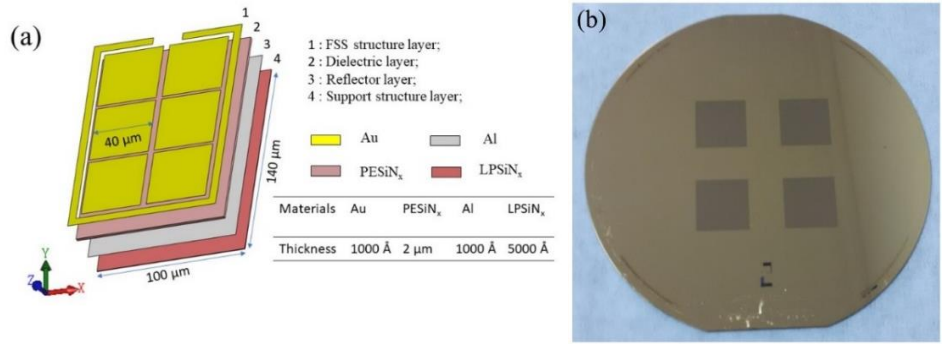

**Fig. 1** Design and fabrication of SQMA: **a** The structure of the SQMA cell. There are four layers in the pixel's cross-sectional view from top to bottom: the 1st layer with 1000 Å Au is used as the FSS structure layer; the 2nd layer with 2μm PESiN$_x$ is the dielectric layer; the 3rd layer with 1000 Å Al is a metal layer acting as a reflector; the 4th layer with 5000 Å LPSiN$_x$ is the support structure. **b** The fabricated 4-inch wafer with SQMA FPA. There are 4 chips on one wafer, and one chip size is 2.5 × 2.5 cm.

Finally, a THz time-domain spectroscopy system was used to measure the FPA absorption for THz radiation; Fig. 2a shows that both the transmitted and reflected THz waves were measured. The THz wave was generated using a low-temperature grown GaAs photoconductive antenna integrated with a Si hemisphere lens. A beam of p-polarized Ti: sapphire femtosecond laser (Maitai Spectraphysics; repetition rate: 80 MHz; pulse width: 70 fs; central wavelength: 800 nm; averaged power: 30 mW) was employed as the excitation source. THz detection was achieved using a ZnTe crystal via electro-optical sampling [35]. The incident angle, θ, of the THz beam was fixed from 20° to 60° (Fig. 2b).

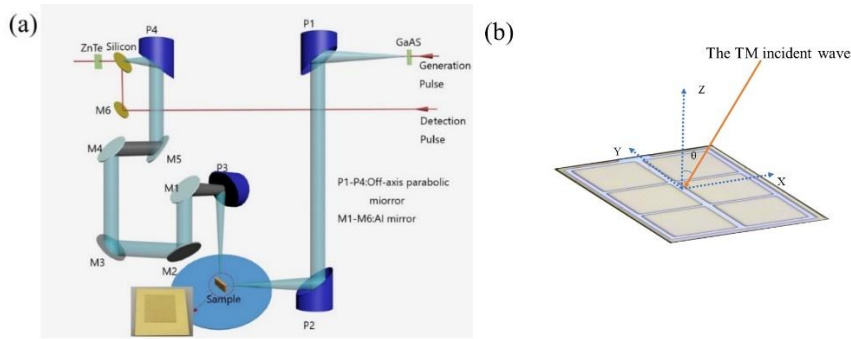

**Fig. 2** THz absorption measurement: **a** Schematic of the THz time-domain spectrometer employed to measure both the reflectance and transmittance. Reflectivity and transmissivity are given by R and T, respectively. Absorptivity of the metamaterial is obtained directly by 1 − R− T. **b** The THz TM incident wave was fixed in the x-z plane, and the incident angle, θ, of the THz beam was fixed from 20° to 60°.

## 3 Results and Quantitative analysis

Fig. 3 shows the absorption-frequency characteristics of SQMA FPA with various incident angles. The absorption characteristics show the characteristics of an absorption peak. The highest absorption peak is 91% at 1.37 THz. The peak value and absorption change with the incident angles and have asymmetric characteristics.

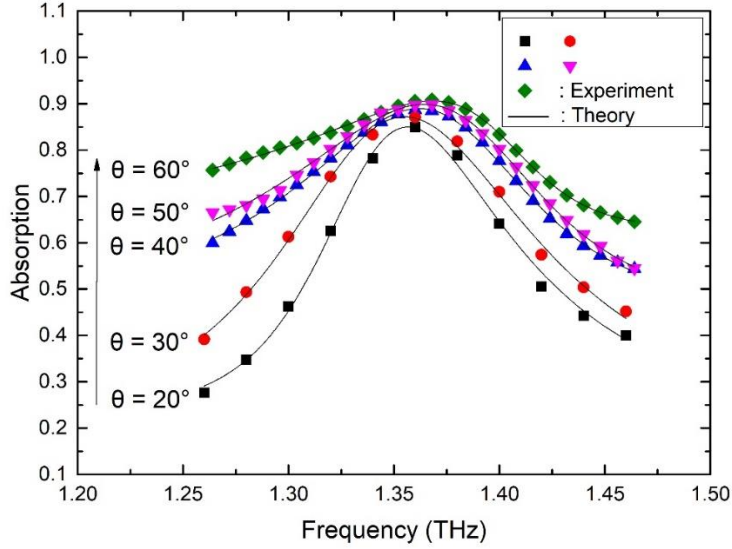

**Fig. 3** Measured absorption of SQMA for TM incident waves with the incident angles from from 20° to 60°. Comparison between the measurement data (scatter) and model (3) (solid line). The curves have asymmetric characteristics.

We constructed an analytical asymmetric model comprising Fermi–Dirac's and Boltzmann's statistics based on probability distributions to explore the absorption characteristics, as follows [36]:

$$p(x) = \frac{\exp((x-x_0)/x_1)}{1+\exp((x-x_0)/x_2)} \quad . \tag{1}$$

The probability (1) is 0.5 at $x_0$; preliminarily, it can be considered A-f proportional to the probability (1):

$$A(f) \propto \frac{A_1 \exp((f-f_0)/f_1)}{1+\exp((f-f_0)/f_2)} \quad . \tag{2}$$

The right side of Equation (2) is a quasi-symmetrical shape; however, A-f is an asymmetric curve. Thus, an additional quantity, $A(f_0) + 0.5A_1$, was added to the right side of Equation (2) to describe the absorption frequency in this study as the peak's asymmetry level.

$$A(f) = \frac{A_1 \exp((f-f_0)/f_1)}{1+\exp((f-f_0)/f_2)} + A(f_0) - \frac{A_1}{2} \quad , \tag{3}$$

where $A_1$ is the maximum amplitude, $A(f_0)$ is the absorption at $f_0$, $f_0$ is the horizontal center position frequency, and $f_1$ and $f_2$ are the shape parameters.

The peak position can be obtained by differentiating Equation (3):

$$\begin{aligned}\frac{d}{df}A(f)\Big|_{f_p} &= 0 = \frac{d}{df}[A_1 \frac{\exp((f-f_0)/f_1)}{\exp((f-f_0)/f_2)+1} + A(f_0) - A_1/2]\Big|_{f_p} \\ &= A_1 \frac{\exp((f-f_0)/f_1)[f_2 - (f_1-f_2)\exp((f-f_0)/f_2)]}{[1+\exp((f-f_0)/f_2)]^2}\Big|_{f_{f_p}}\end{aligned} \tag{4}$$

The peak position is given as follows：

$$f_p = f_0 + f_2 Ln \frac{f_2}{f_1 - f_2}. \tag{5}$$

Model (3) was fit to the SQMA raw data and pooled to form an overall absorption-frequency relationship that obtains information from each absorption (Fig. 3). The best-fitting model parameters $A_1$, $A(f_0)$, $f_0$, $f_1$, and $f_2$ of model (3) were anticipated using the least-squares nonlinear curve regression analysis method (Table 1). Model (3) was evaluated by $R^2$ as the coefficient of determination. This is a simple and efficient confidence level to evaluate the model fit. Using $R^2$ to evaluate the fitting emphasizes the accuracy of the prediction using the model curve. The peak position, $f_p$, was obtained from Equation (5), as shown in Table 1. There was a difference between $f_0$ and $f_p$ in third place after the decimal point. This simply shows that the absorption curve is asymmetric.

To verify the universality of model (3), parameter estimation from the third part of the experimental data is critical to achieving the desired model predictive properties. The raw data used in this study were adapted from the paper: fishnet-like MIM-based film terahertz absorbers [8]. There are ten absorption curves of fishnet-like absorbers at different incidence angles [8]. There are ten absorption curves of fishnet-like absorbers at different incidence angles. One major objective of this study is to figure out what physical features are responsible for THz metamaterial absorption. Such many absorption curves are sufficient for determining the physical features of THz metamaterial absorption. Model (3) will be fit to the raw data of MIM absorbers and pooled to form an overall absorption-frequency relationship that obtains information from each absorption curve (Fig. 4). Since model (3) satisfactorily reflects the actual THz absorption behavior of MMAs, each parameter of the model was applied to identify the TE and TM incidence angles for fishnet-like absorbers. The best-fitting model parameters $A_1$, $A(f_0)$, $f_0$, $f_1$, and $f_2$ of model (3) are anticipated using the least-squares nonlinear curve regression analysis method (Table 1). The minimum $R^2$ is greater than 0.99. The confidence level of each parameter is over 99%.

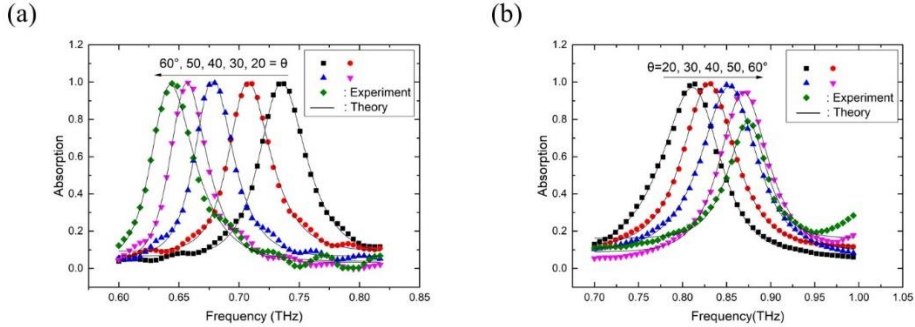

**Fig. 4** Comparison between the raw data (scatter) and fitting in model (3) (solid line): **a** fishnet-like absorber with TM wave; **b** fishnet-like absorber with TE wave. The raw data were adapted from the paper: fishnet-like MIM-based film terahertz absorbers [8].

**Table 1 Optimal absorption-frequency parameters at different incident angles**

| MMAs | polarization | θ(°) | $A_1$ | $f_0$(THz) | $f_1$(THz) | $f_2$(THz) | $A(f_0)$ | $f_p$(THz) | $R^2$ |
|---|---|---|---|---|---|---|---|---|---|
|  |  | 20 | 1.19 | 1.344 | 0.0288 | 0.019 | 0.81778 | 1.357 | 0.992 |
| SQMA |  | 30 | 1.24 | 1.345 | 0.04236 | 0.02528 | 0.860 | 1.355 | 0.993 |
|  | TM | 40 | 0.81 | 1.375 | 0.063 | 0.023 | 0.875 | 1.362 | 0.998 |

| | | 50 | 0.78 | 1.385 | 0.0813 | 0.024 | 0.8638 | 1.364 | 0.996 |
| | | 60 | 0.43 | 1.399 | 0.115 | 0.0175 | 0.8415 | 1.369 | 0.999 |
| Fishnet-like | TM | 20 | 1.76 | 0.8157 | 0.03403 | 0.01433 | 0.959745 | 0.811 | 0.997 |
| | | 30 | 1.74 | 0.8298 | 0.02584 | 0.01302 | 0.99036 | 0.830 | 0.999 |
| | | 40 | 1.75 | 0.8548 | 0.02933 | 0.01374 | 0.974795 | 0.853 | 0.998 |
| | | 50 | 1.71 | 0.8673 | 0.02241 | 0.01205 | 0.94299 | 0.869 | 0.992 |
| | | 60 | 1.24 | 0.8749 | 0.02044 | 0.00987 | 0.78329 | 0.874 | 0.996 |
| Fishnet-like | TE | 20 | 1.84 | 0.7328 | 0.01476 | 0.00846 | 0.985225 | 0.735 | 0.996 |
| | | 30 | 1.78 | 0.7052 | 0.0125 | 0.00748 | 0.984635 | 0.708 | 0.996 |
| | | 40 | 1.88 | 0.6784 | 0.01229 | 0.00713 | 1.003425 | 0.681 | 0.995 |
| | | 50 | 1.88 | 0.6574 | 0.01157 | 0.00704 | 0.97499 | 0.661 | 0.992 |
| | | 60 | 1.81 | 0.6397 | 0.01205 | 0.00778 | 0.945705 | 0.644 | 0.991 |

$R^2$: Coefficient of Determination;

## 4 Quadratic model analysis

Model (3) can be used to predict other experimental conditions that were not tested. Table 1 shows that these parameters and angles have certain regularity. We developed a quadratic model analysis of the coefficients of the asymmetric model. Fig. 5–7 show the curve coefficients $f_0$, $f_1$, $f_2$, and $A(f_0)$ of the absorption model versus incident angles.

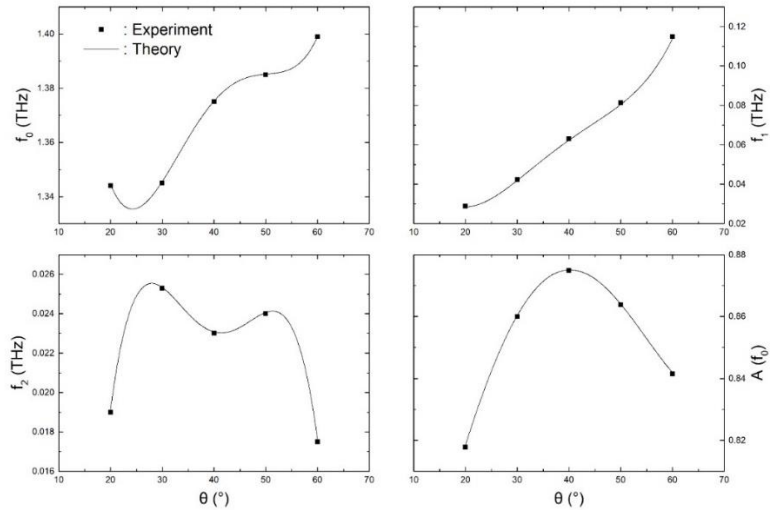

**Fig. 5** Parameters of the fitted $f_0$, $f_1$, $f_2$, and $A(f_0)$ results compared with those listed in Table 1. The absorption model of SQMA with TM incident wave. (Scatter line: listed in Table 1; solid line: fitting)

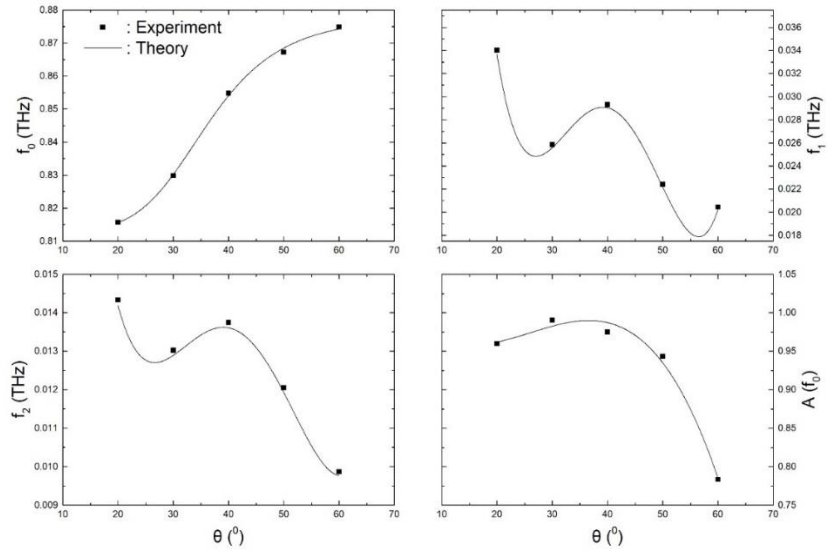

**Fig. 6** Parameters of the fitted $f_0$, $f_1$, $f_2$, and $A(f_0)$ results compared with those listed in Table 1. The absorption model of fishnet-like with TM incident wave. (scatter line: listed in Table 1; solid line: fitting)

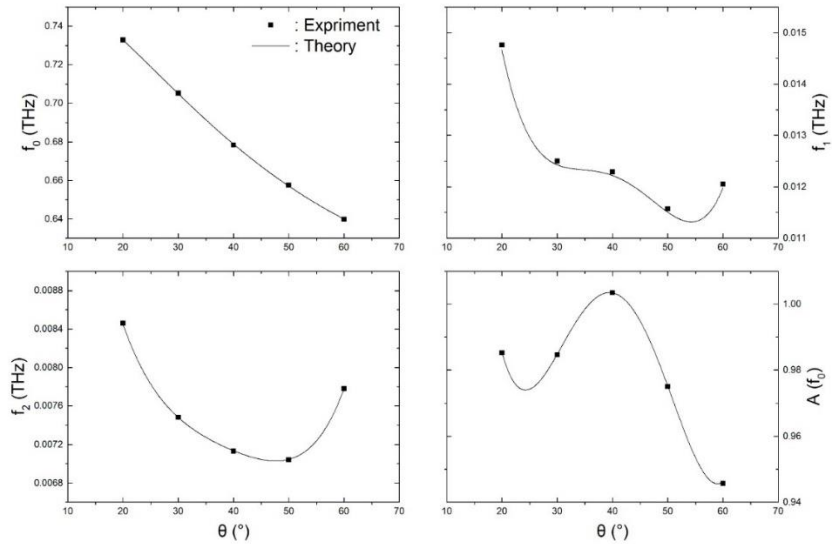

**Fig. 7** Parameters of the fitted $f_0$, $f_1$, $f_2$, and $A(f_0)$ results compared with those listed in Table 1. The absorption model of fishnet-like with TE incident wave. (scatter line: listed in Table 1; solid line: fitting)

Furthermore, as shown in Figs. 5–7, the increase in the angles results in different variations in the absorption-frequency model parameters ($f_0$, $f_1$, $f_2$, $A(f_0)$). The model (3) fitted very well with the experiment data. Fig. 5 shows that the fitting results of SQMA with the TM incidence wave. Fig. 6 shows that the fitting results of fishnet-like with the TE incidence wave. Fig. 7 shows that the fitting results of fishnet-like with the TM incidence wave. This reflects the accuracy of the secondary model.

Several stepwise regression analyses were performed using the data in Table 1 to obtain the mathematical relationship between the basic angle and model (3) parameters. The quantitative relationship between the incidence angle and $f_0$, $f_1$, $f_2$, and $A(f_0)$ was provided from expression (6) – (17) in line with Origin9.1. Expressions (10 and 14) have the logistic model law. Expression (13) has the law of 3$^{rd}$ power polynomial law. The other formulas have the 4$^{th}$ power polynomial law. It was intended that approximate expressions in assessing the parameters exclusively from θ could be achieved by conducting the process, which could eventually result in the following expressions:

SQMA with TM incident wave:

$$f_0(\theta) = 1.9841 - 0.07447\theta + 0.00301\theta^2 - 5.02425 \times 10^{-5}\theta^3 + 3.01125 \times 10^{-7}\theta^4. \quad (6)$$

$$f_1(\theta) = 0.19444 - 0.0217\theta + 9.7 \times 10^{-4}\theta^2 - 1.7324 \times 10^{-5}\theta^3 + 1.12 \times 10^{-7}\theta^4. \quad (7)$$

$$f_2(\theta) = -0.1776 + 0.02219\theta - 8.8 \times 10^{-4}\theta^2 + 1.5 \times 10^{-5}\theta^3 - 9.33 \times 10^{-8}\theta^4. \quad (8)$$

$$A(\theta, f_0) = 0.7175 + 0.00265\theta + 2.53 \times 10^{-4}\theta^2 - 7.81 \times 10^{-6}\theta^3 + 5.717 \times 10^{-8}\theta^4. \quad (9)$$

Fishnet-like absorbers with TM incident wave:

$$f_0(\theta) = 0.878 + \frac{0.813 - 0.878}{1 + (\theta/36.3)^{5.43}}. \quad (10)$$

$$f_1(\theta) = 0.3611 - 0.0373\theta + 0.0015\theta^2 - 2.5528 \times 10^{-5}\theta^3 + 1.56 \times 10^{-7}\theta^4. \quad (11)$$

$$f_2(\theta) = 0.072 - 0.0066\theta + 2.622 \times 10^{-4}\theta^2 - 4.4 \times 10^{-6}\theta^3 + 2.62 \times 10^{-8}\theta^4. \quad (12)$$

$$A(\theta, f_0) = 1.037 - 0.0117\theta + 5.34 \times 10^{-4}\theta^2 - 6.81 \times 10^{-6}\theta^3. \quad (13)$$

Fishnet-like absorbers with TE incident wave:

$$f_0(\theta) = 0.56992 + \frac{0.76467 - 0.56992}{1 + (\theta/45)^2}. \quad (14)$$

$$f_1(\theta) = 0.0565 - 0.00454\theta + 1.73221 \times 10^{-4}\theta^2 - 2.88832 \times 10^{-6}\theta^3 + 1.76137 \times 10^{-8}\theta^4. \quad (15)$$

$$f_2(\theta) = 0.01838 - 0.001\theta + 3.614 \times 10^{-5}\theta^2 - 6.039 \times 10^{-7}\theta^3 + 3.8775 \times 10^{-9}\theta^4. \quad (16)$$

$$A(\theta, f_0) = 1.86711 - 0.1052\theta + 0.00439\theta^2 - 7.62358 \times 10^{-5}\theta^3 + 4.66042 \times 10^{-7}\theta^4. \quad (17)$$

Figs. 8–10 show that the secondary model is derived from the asymmetric model and predicts the absorption spectrum for other angles of incidence. In Fig. 8, the SQMA absorption curve has the shape of a "valley" at incident angles of 5°, 70°, and 80°. In Fig. 9, the absorption curve becomes narrow at an incident angle of 10°. In Fig. 10, the valley curve occurs at incident angles of 70° and 80°. A critical point angle of 66° from the "peak" to the "valley" is investigated in depth. The red color line is used to plot the particular curve. At a particular incident angle, the device has transitioned from an absorber to a filter. This method may provide ideas and methods for discovering new physical phenomena and promoting more in-depth physical analysis.

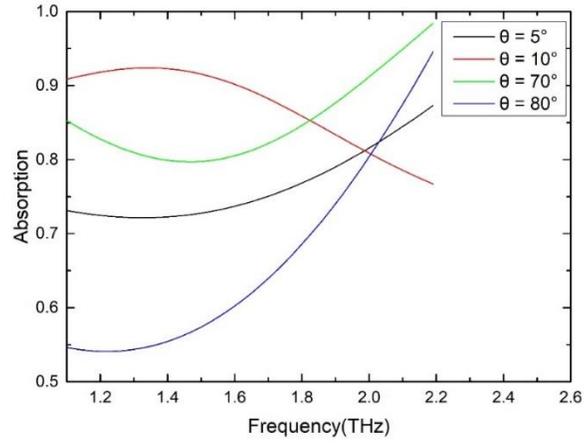

**Fig. 8** Prediction curve of SQMA absorber for different angles of TM wave incidence.

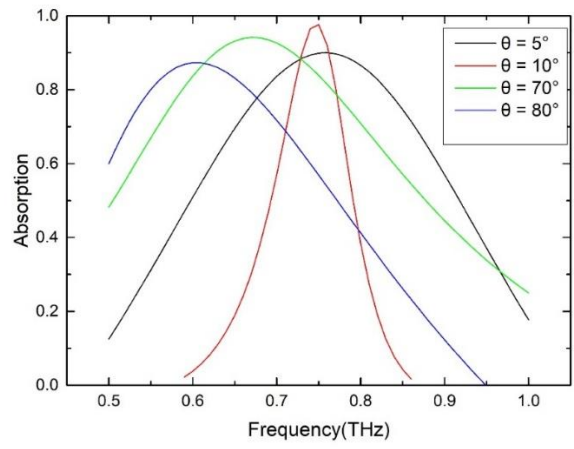

**Fig. 9** Prediction curve of fishnet-like absorber for different angles of TM wave incidence

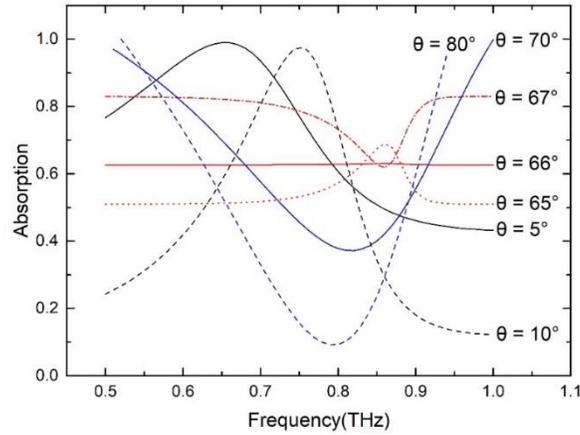

**Fig. 10** Prediction curve of fishnet-like absorbers for different angles of TE wave incidence. The critical point angle 65°,66°, 65° are drawn in red lines.

## 5 Conclusion

In this study, a MIM perfect absorber focal plane array was designed, fabricated, and characterized. An analytical model for extracting the absorption-frequency characteristics of MMAs during the THz detection process was proposed. Unlike traditional absorber models, asymmetric absorption behavior at different incidence angles was analyzed using the revised Fermi–Dirac's and Boltzmann's statistics. The validation of the asymmetric model was consistent with the experimental data from the characterized results and third-party research. A secondary model of this model was established by analyzing the changing law of model parameters. From the secondary model analysis, $3^{rd}$-order, $4^{th}$-order polynomial, and logistic laws were discovered in the model parameters. Using the secondary model, we successfully predicted the absorption characteristics of the absorbers for other incident angles that were not tested. Some novel phenomena were discovered in the analysis. The absorption curve changed from peak to valley at certain incident angles. The MIM absorber asymmetric model and entity validation showed that the proposed model can accurately acquire absorption over the THz frequency for THz detection. It is expected that this model method can be used to explore new THz wave absorption laws and new phenomena of interaction between THz waves and metamaterials. This model can also be applied to the THz band and visible, microwave, infrared, and other frequency bands.

**Acknowledgements**

This work is supported by the National Natural Science Foundation of China with Grant NO. 61874137.

**Data availability statements**
All data generated or analysed during this study are included in this published article [and its supplementary information files].

# References

1. P. Yu, L. V. Besteiro, Y. Huang, J. Wu, L. Fu, H. H. Tan, C. Jagadish, G. P. Wiederrecht, A. O. Govorov, and Z. Wang, "Broadband Metamaterial Absorbers," Advanced Optical Materials **7**, 1800995 (2019).
2. C. M. Watts, X. Liu, and W. J. Padilla, "Metamaterial Electromagnetic Wave Absorbers," Advanced Materials **24**, OP98-OP120 (2012).
3. H. T. Yudistira and K. Kananda, "Design of wideband single-layer metamaterial absorber in the S-band and C-band spectrum," The European Physical Journal Plus **136**, 603 (2021).
4. H. Wang, B. Yan, H. Jin, Z. Wang, L. Guo, B. Li, B. Yu, and C. Gong, "Perfect absorber with separated 'dielectric–metal–ground' metamaterial structure," Journal of Physics D: Applied Physics **54**, 225105 (2021).
5. Z. Shen, S. Li, Y. Xu, W. Yin, L. Zhang, and X. Chen, "Three-Dimensional Printed Ultrabroadband Terahertz Metamaterial Absorbers," Physical Review Applied **16**, 014066 (2021).
6. M. Pourmand, P. K. Choudhury, and M. A. Mohamed, "Tunable absorber embedded with GST mediums and trilayer graphene strip microheaters," Scientific Reports **11**, 3603 (2021).
7. L. Huang, D. R. Chowdhury, S. Ramani, M. T. Reiten, S.-N. Luo, A. J. Taylor, and H.-T. Chen, "Experimental demonstration of terahertz metamaterial absorbers with a broad and flat high absorption band," Opt. Lett. **37**, 154-156 (2012).
8. D. Y. Shchegolkov, A. K. Azad, J. F. O'Hara, and E. I. Simakov, "Perfect subwavelength fishnetlike metamaterial-based film terahertz absorbers," Physical Review B **82**, 205117 (2010).
9. Y. Wu, H. Lin, J. Xiong, J. Hou, R. Zhou, F. Deng, and R. Tang, "A broadband metamaterial absorber design using characteristic modes analysis," Journal of Applied Physics **129**, 134902 (2021).
10. V. G. Veselago, "The electrodynamics of substances with simultaneously negative values of ε and μ," Soviet Physics Uspekhi **10**, 509-514 (1968).
11. J. B. Pendry, "Negative Refraction Makes a Perfect Lens," Physical Review Letters **85**, 3966-3969 (2000).
12. D. R. Smith, W. J. Padilla, D. C. Vier, S. C. Nemat-Nasser, and S. Schultz, "Composite Medium with Simultaneously Negative Permeability and Permittivity," Physical Review Letters **84**, 4184-4187 (2000).
13. H. Tao, N. I. Landy, C. M. Bingham, X. Zhang, R. D. Averitt, and W. J. Padilla, "A metamaterial absorber for the terahertz regime: design, fabrication and characterization," Opt Express **16**, 7181-7188 (2008).
14. N. Liu, M. Mesch, T. Weiss, M. Hentschel, and H. Giessen, "Infrared Perfect Absorber and Its Application As Plasmonic Sensor," Nano Letters **10**, 2342-2348 (2010).
15. N. I. Landy, C. M. Bingham, T. Tyler, N. Jokerst, D. R. Smith, and W. J. Padilla, "Design, theory, and measurement of a polarization-insensitive absorber for terahertz imaging," Physical Review B **79**, 125104 (2009).
16. Z. Sakotic, A. Krasnok, N. Cselyuszka, N. Jankovic, and A. Alú, "Berreman Embedded Eigenstates for Narrow-Band Absorption and Thermal Emission," Physical Review Applied **13**, 064073 (2020).
17. R. Mittra, C. H. Chan, and T. Cwik, "Techniques for analyzing frequency selective surfaces-a review," Proceedings of the IEEE **76**, 1593-1615 (1988).
18. Y. Ra'di, C. R. Simovski, and S. A. Tretyakov, "Thin Perfect Absorbers for Electromagnetic Waves: Theory, Design, and Realizations," Physical Review Applied **3**, 037001 (2015).
19. T. Q. H. Nguyen, T. K. T. Nguyen, T. N. Cao, H. Nguyen, and L. G. Bach, "Numerical study of a broadband metamaterial absorber using a single split circle ring and lumped resistors for X-band applications," AIP Advances **10**, 035326 (2020).
20. H. T. Chen, "Interference theory of metamaterial perfect absorbers," Opt Express **20**, 7165-7172 (2012).
21. M. Mohammadi, H. Rajabalipanah, and A. Abdolali, "A theoretical investigation on reciprocity-inspired wide-angle spectrally-selective THz absorbers augmented by anisotropic metamaterials," Scientific Reports **10**, 10396 (2020).
22. O. Luukkonen, C. Simovski, G. Granet, G. Goussetis, D. Lioubtchenko, A. V. Raisanen, and S. A. Tretyakov, "Simple and Accurate Analytical Model of Planar Grids and High-Impedance Surfaces Comprising Metal Strips or Patches," IEEE Transactions on Antennas and Propagation **56**, 1624-1632 (2008).
23. W. Wang, F. Yan, S. Tan, H. Li, X. Du, L. Zhang, Z. Bai, D. Cheng, H. Zhou, and Y. Hou, "Enhancing sensing capacity of terahertz metamaterial absorbers with a surface-relief design," Photon. Res. **8**, 519-527 (2020).
24. H. Luo, Y. Z. Cheng, and R. Z. Gong, "Numerical study of metamaterial absorber and extending absorbance bandwidth based on multi-square patches," The European Physical Journal B **81**, 387-392 (2011).
25. S. Wang, C. Cai, M. You, F. Liu, M. Wu, S. Li, H. Bao, L. Kang, and D. H. Werner, "Vanadium dioxide based broadband THz metamaterial absorbers with high tunability: simulation study," Opt. Express **27**, 19436-19447 (2019).
26. K. Tantiwanichapan and H. Durmaz, "Herbicide/pesticide sensing with metamaterial absorber in THz regime," Sensors and Actuators A: Physical **331**, 112960 (2021).
27. W. Pan, T. Shen, Y. Ma, Z. Zhang, H. Yang, X. Wang, X. Zhang, Y. Li, and L. Yang, "Dual-band and polarization-independent metamaterial terahertz narrowband absorber," Appl. Opt. **60**, 2235-2241 (2021).
28. F. Alves, L. Pimental, D. Grbovic, and G. Karunasiri, "MEMS terahertz-to-infrared band converter using frequency selective planar metamaterial," Sci Rep **8**, 12466 (2018).
29. J. Bai, Z. Pang, P. Shen, T. Chen, W. Shen, S. Wang, and S. Chang, "A terahertz photo-thermoelectric detector based on metamaterial absorber," Optics Communications **497**, 127184 (2021).
30. H. Pan and H. Zhang, "Thermally tunable polarization-insensitive ultra-broadband terahertz metamaterial absorber based on the coupled toroidal dipole modes," Opt. Express **29**, 18081-18094 (2021).

31. J. Li, J. Li, H. Zhou, G. Zhang, H. Liu, S. Wang, and F. Yi, "Plasmonic metamaterial absorbers with strong coupling effects for small pixel infrared detectors," Opt. Express **29**, 22907-22921 (2021).
32. R. Xu, X. Liu, and Y.-S. Lin, "Tunable ultra-narrowband terahertz perfect absorber by using metal-insulator-metal microstructures," Results in Physics **13**, 102176 (2019).
33. Z. Ren, R. Liu, H. Lu, Y. Guo, and R. Xie, "Tunable ultranarrow-band metamaterial perfect absorber based on electromagnetically induced transparency structure," Optical Materials **122**, 111624 (2021).
34. G. Yoon, I. Kim, and J. Rho, "Challenges in fabrication towards realization of practical metamaterials," Microelectronic Engineering **163**, 7-20 (2016).
35. Y. Zhou, Y. E, L. Zhu, M. Qi, X. Xu, J. Bai, Z. Ren, and L. Wang, "Terahertz wave reflection impedance matching properties of graphene layers at oblique incidence," Carbon **96**, 1129-1137 (2016).
36. P. Zhang, Y. Feng, X. Wen, W. Cao, R. Anthony, U. Kortshagen, G. Conibeer, and S. Huang, "Generation of hot carrier population in colloidal silicon quantum dots for high-efficiency photovoltaics," Solar Energy Materials and Solar Cells **145**, 391-396 (2016).